\documentclass[10pt]{article}
\usepackage{epsfig}
\usepackage{graphics}
\usepackage[english]{babel}      

\newcommand{\be}{\begin{equation}}
\newcommand{\ee}{\end{equation}}
\newcommand{\ba}{\begin{eqnarray}}
\newcommand{\ea}{\end{eqnarray}}

\begin{document}

\title{On the Deconfinement Phase Transition in Heavy-Ion Collisions}

\author{Abdelnasser M. Tawfik\\ \\
			{\normalsize\sf FB Physik, Marburg University, Renthof 5, D-35032 Marburg, Germany} 
}

\date{}
\maketitle
\parindent5mm

\renewcommand{\thefootnote}{\arabic{footnote}}

\begin{abstract}

{The factorial moments (FM) of multiplicity distribution 
are used to study the deconfinement
phase transition in heavy-ion collisions. The relation between
FM and the partition number, $M$, results positive intermittency
exponents, $\phi_q$. 
According to the signatures suggested from certain statistical models, 
the two-dimensional results of the dependence of $\phi_q/\phi_2$, anomalous fractal, 
$d_q/d_2$ and R\'enyi dimensions, ${\cal R}_q/{\cal R}$, and the
normalized exponents, $\zeta_q$, on the orders of FM 
evidently supply evidence for the
quark-hadron phase transition in Pb+Pb collisions at 158 AGeV.} \\

Quark-gluon plasma 
\end{abstract}

\section{Introduction}
\label{sec1}

The non-statistical fluctuations in the final state of particle
production are suggested as an experimental tool to diagnostically
confirm the elusive quark-hadron phase transition 
\cite{GQP1,GQP2,GQP3,GQP4}.
Ever since the observations of spike-events first observed in the
cosmic ray interactions \cite{jacee} and later re-produced in the
laboratory \cite{spik} and since the pioneer works of Bia{\l}as
and Peschanski \cite{Bia86}, the intermittent behavior has
attracted a lot of attention. Therefore, it has been examined in
different interacting systems. If the quark-hadron phase
transition is to be produced in heavy-ion collisions, its
critical aftereffects (like {\it soft} and {\it hard} collisions,
clustering and resonance decay, showering processes, critical
exponents, etc.) are expected to survive even until the {\it
freeze-out} \cite{Rajag}. They reflect themselves in form of non-statistical
(e.g. dynamical) fluctuations, power-scaling behavior,
self-similar branching, mono-fractal density fluctuations, etc. In this
letter, we study FM of the multiplicity distribution through 
successive partitions in one- and in two-dimensions (pseudo-rapidity, $\eta$, and/or
azimuthal angle, $\phi$). Using these investigations, we shall try to access the
deconfinement phase transition \cite{GQP4}. For this destination,
we will utilize the attitude of intermittency exponents, anomalous fractal, and
R\'enyi dimensions on the orders of FM. QGP signatures principally 
suggested according to certain statistical models will be read off by means of 
the relation between intermittency and multi-fractality. \\

The data sample used for this work is retrieved from some of the
Pb-chambers irradiated at CERN-SPS during 1996 for the EMU01
collaboration. Specifically, the data sample employed here is
that we have completely analyzed by our automatic measuring 
system MIRACLE Lab \cite{taw12} at Marburg university. 
For details about the emulsion irradiation, we refer to
\cite{taw-jpc}. Due to their high resolution and geometrical
acceptance, the nuclear emulsions are effectively capable to measure the
charged particles, the angular distribution, and the density
fluctuations of high multiplicity collisions. The scanning
efficiency in emulsion chambers is $\sim 0.75\pm 0.05$
\cite{EMU01-1}. The sensitivity for singly charged particles is
as good as 30 grains per 100 $\mu$m. It is therefore, possible to
locate the track positions to an accuracy $<2~\mu$m.
Depending on the topology of the microscope's field of view, 
the produced particles with space angles, $\theta<
30^{\circ}$ (pseudorapidity, $\eta=-\ln\tan(\theta/2)>1.32$) 
can be acquired.

The produced particles are expected to be mixed with
contamination of electron-pairs from Dalitz decays and undetected 
$\gamma$-conversions. The possible {\it overestimation} of 
particle density has been determined as $\sim 2\%$ \cite{TawDis}.
As a reason of the reconstruction algorithm applied for
MIRACLE Lab, the tracks of these electrons are {\it
completely} disregarded or percolated.
The efficiency of MIRACLE Lab is estimated as $\sim 96\%$. 
From manual/automatic 
comparison we noticed that the automatic reconstruction 
underestimates the multiplicity, especially the particles with 
relatively wide angles ($\eta<2$). Besides these missing 
measurements, the frequent scattering, unresolved 
close-pairs, nearest-neighbor particles, and pair production represent an additional 
source of this $4\%$-discrepancy \cite{taw12}. Many of the extra tracks in the 
manual measurement are close pairs separated by $< 1~\mu$m. 
The two track resolution, close neighbors, and the split track 
recognition are discussed in \cite{taw12}.  
In the automatic measuring facility, the lake of close-pairs within $< 1~\mu$m can be
referred to the limited instrumental pair resolution. Also 
if the particle trajectory is spilt into two tracks, which can result a 
huge intermittency signal, both of them are rejected, since  their 
own vertex definitely will not be coincident with the common one \cite{taw12}.   
Then we can summarize that the data sample grabbed by MIRACLE Lab is
evidently liberated from the double counting, measuring bias, and from any
contaminations from particle decays or secondary interactions. We
could therefore, consciously suggest to disregard 
the effects of Dalitz decays and $\gamma$-conversions on FM
\cite{taw-jpc} ({\it see} Sect.~\ref{BEK-FM}  below). 

The accuracy of measured $\eta$ and $\phi$ is practically depending on  
determination of the event axis and on the emulsion plates, into which the 
corresponding track penetrates until it leaves the field of view. 
The $x$- and $y$-coordinates are ideally to be measured with 
respect to some reference points whose positions are well-determined.  
The location of track in such transverse plane has a statistical 
uncertainty of $\sim 0.3~\mu$m. On the average, the tracks 
leave the field of view at distance \hbox{$\sim 75~\mu$m} from the event exis. 
Then the resulting uncertainty in $\eta$ is $\sim 0.004$. 
Due to small air gaps, non-uniformity of the spacer between the plates, variations in 
emulsion thickness,  the $z$-coordinates are also uncertain. 
For most of tracks, the last measured emulsion plate is located $2.5~$cm downstream. 
The uncertainty in $\phi$ is then $\sim 0.136~${\it mrad} \cite{TawDis}. 

\setlength{\textfloatsep}{10pt}
\begin{figure}[htb]
\begin{center}{\epsfxsize=6.5cm \epsfbox{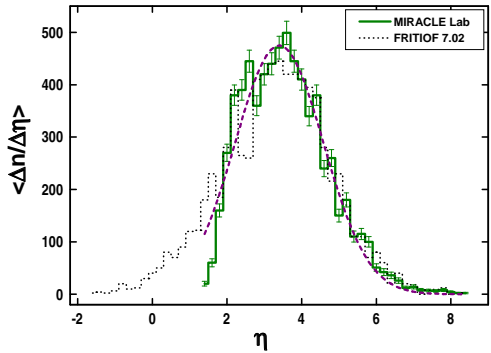}}  
 \parbox{9cm}{\caption{\small\it Pseudorapidity distribution of central Pb+Pb events measured 
by MIRACLE Lab (solid line). The dotted line represents the distribution of FRITIOF, meanwhile
the dashed line gives the Gaussian fit. The FRITIOF sample is simulated with 
zero impact parameter and default configurations.
 \label{fig:qgp0}}}
\end{center}
\vspace{5pt} 
\end{figure}

Fig.~\ref{fig:qgp0} shows the pseudorapidity distribution of Pb+Pb central 
events with multiplicity $\ge 1200$. In order to compare the experimental data to 
the expectations based on incoherent models, we have simulated a sample 
of Pb+Pb collisions using FRITIOF 7.02 Monte Carlo code 
with very small impact parameters. FRTITIOF code 
has been run in its default configuration. The dotted Line shows the 
pseudorapidity distribution of such sample. For $\eta$ grater than $\sim 1.8$, 
the two distributions are in 
good agreement. There is a small flattening in the central 
peak of experimental distribution. The flattening is expected, if 
QGP had been produced \cite{bjork}. \\

If a phase-space of width $\Delta$ is split into $M$ equal bins of
size, \hbox{$\delta=\Delta/M$}, the scaled factorial
moments of multiplicity distribution are defined \cite{Bia86} as:
\be
F_q(M)= M^{q-1} \sum\limits_{m=1}^{M}
       \frac{\left<n_m(n_m-1)\cdots(n_m-q+1)\right>}{\bar{n}^q},
\label{Eq:1}
\ee
$n_m$ is the multiplicity in $m$-th bin and $\bar{n}$ is the
average multiplicity in the whole $\Delta$-window. According to
the self-similar density fluctuations \cite{Hwa-Self} (e.g. QCD parton
cascading), the successive partitions lead to the following
power-law \cite{Bia86,Chaos1}:

\be
F_q(M) \;\; \propto \;\; M^{\phi_q}.
\label{Eq:2}
\ee

The term {\it ''intermittency''} used for the turbulence in the theory of chaos
to describe the development of a hydro-dynamical system from
stable to chaotic state can analogously be used in the particle
production \cite{Chaos1,Chaos2}. The exponents, $\phi_q$, called {\it
''intermittency exponents''}, can be determined 
from the asymptotic behavior. Their behavior elevates the 
interpretation of the intermittency as aftereffects of 
quark-hadron phase transition \cite{GQP2,Bial11,Bial12,Bial13}. Based 
on 2D Ising model \cite{satz21,satz22,satz1}, the same 
interpretation is obtained. In addition to these statistical results, 
it is proposed that the intermittency parameters
possibly {\it contain} signatures for the phase transition in heavy-ion 
collisions \cite{GQP2,GQP3}. In the contrast, assuming that the self-similarity 
dominates the particle production, it was difficult to {\it concretely} suggest 
their intermittency as signature of QGP \cite{GQP3,Hwa-Self}. 
This present work distinguishes between all these scenarios. \\

$\phi_q$ are related to the {\it ''anomalous fractal dimensions''} 
though the following relation \cite{Hent}:
\be
d_q = \phi_q\cdot(q-1)^{-1}.
\label{Eq:3}
\ee
At the critical point of Ising model \cite{satz21,satz1}, 
$d_q$ can be given in terms of critical exponents. 
Also at this point, $d_q$ are independent on the orders, $q$. Generally, 
the $q$-dependency of $d_q$ interdepends on the nature of processes taking place
in the interacting system. Therefore, $d_q$ are liable to the branching processes that 
precipitate an intermittent behavior. For this reason, $d_q$ can be successful candidates to give 
a further signature for the QGP formation \cite{GQP3}.\\

The {\it fractal R\'enyi dimensions}, which ordinarily, are used to measure the
randomization in particle production, are, in turn, depending on
$d_q$ \cite{GQP2,Reny1,Reny2},
\be {\cal R}_q = {\cal R} \cdot (1-d_q), 
\label{Eq:4}
\ee
The constant, ${\cal R}$, represents the topological R\'enyi
dimension. For multi-fractal processes, $d_q$ are linearly depending
on $q$. For mono-fractal density fluctuations, $d_q$ are
constant. \\

Assuming that the intermittency and the mono-fractal density fluctuations 
survive the further phases until the freeze-out \cite{Rajag}, 
different proposals are suggested to indicate the deconfinement phase transition 
via the interplay between intermittency and mono-fractality. 
From statistical models, it is expected 
that at $T_c$ the intermittency are related to 
only one particular combination of the critical exponents. In Eq.~\ref{Eq:3}, 
$\phi_q$  have been squeezed out in $d_q$ and then in ${\cal R}_q$ (Eq.~\ref{Eq:4}).
Hence, the parameter governing the intermittency at a thermal phase 
transition can be represented by constant ${\cal R}_q$. 
Ordinarily, it is supposed that the mono-fractality measures
the random sets of intermittent clusters of the {\it confined} phase 
inside the {\it deconfined} one \cite{Opall,Reny3}. 
In a second-order QCD phase transition, it is expected that intermittent fluctuations   
take place and therefore, are characterized by unique anomalous 
fractal dimension. In 2D Ising model, the same behavior 
with constant $d_q$ has been proven \cite{satz21,satz1,Opall}. 
The validity of Eq.~\ref{Eq:8} is proposed as a signature of QGP
\cite{GQP2}, i.e. the validity of $d_q/d_2=1$ 
according to Eq.~\ref{Eq:5}. The expectations given in this section represent 
the model used in this letter to study the quark-hadron 
phase transition.

\section{Analysis and Results}
\subsection{Scaled Factorial Moments}
\label{sec2}

\setlength{\textfloatsep}{10pt}
\begin{figure}[htb]
\begin{center}{\epsfxsize=6.5cm \epsfbox{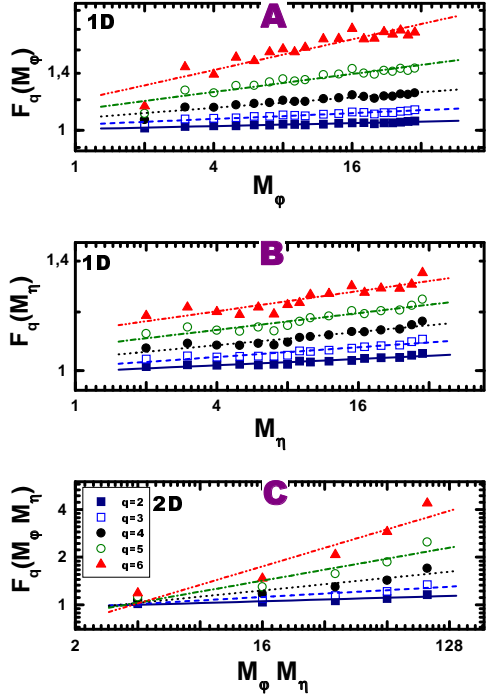}}  
 \parbox{9cm}{\caption{\small\it The $q$-order FM are given as functions of the
  partition number, $M$, and drawn in a log-log scale. In the top picture
  ({\bf A}) the partition is performed in $\phi-$dimension only. The results of the partition in
  $\eta-$dimension are given in picture {\bf B}. The two-dimensional partition
  ($\eta$ and $\phi$) is illustrated in the bottom part ({\bf C}).
  We notice that the interacting system is obviously intermittent in one- as well as
  in two-dimensions.
 \label{fig:qgp1}}}
\end{center}
\vspace{5pt} 
\end{figure}

FM are calculated for the Pb+Pb events with the highest particle multiplicity ($\ge 1200$). 
The events with lower multiplicities are sampled, incompletely. 
The smallest one has a multiplicity of 410. As a result of this selection process, 
events at a rate $\sim 0.002$ per incoming primary beam are chosen for FM. 
Based on parameterization of 
charge-changing cross-section for heavy-ion interactions,
we obtain $\sigma_{Pb+Pb}\approx 6.71\pm 0.21~$barn \cite{Cross-S}.
According to the distribution shape in the forward cone, 
the centrality of analyzed events can be estimated. 
The central ones contain two or fewer projectile 
fragmentations and have the highest multiplicity. 
The other events are classified as peripheral or semicentral. 
Furthermore, the {\it central}
events are characterized by total break-up into singly
charged particles \cite{evert} and  
supposed to be more favorable to produce QGP, 
as a result of the more nuclear matter they include.
Besides the definition of interaction {\it centrality}, 
the multiplicity restriction can be employed to avoid any {\it possible
mechanical} correlations \cite{taw-jpc} and to determine the kind of FM \cite{TawDis}. 
From these events, only the particles
emitted within a predefined $\eta$-interval,
\hbox{$2<\eta<6$}, are taken into account. This interval
obviously contains the region of central rapidity. Due to the
chiral symmetry breaking, the produced particles are mainly
pions. Therefore, the restriction on $\eta$ is significant
to study the deconfinement phase transition, which, in turn, is
restricted to the produced particles. Here we employ
an additional restriction on the considered phase-space.
Its influence to get {\it flat} FM distribution
has been discussed elsewhere \cite{taw-jpc}. On the azimuthal space,
there is no restriction, i.e. for each $\Delta\eta$ group, 
$\delta\phi$-bins are allowed to take any real value within the
available spectrum $\{0,2\pi\}$. \\

From the analyzed $380~$events, 84 central events are selected according the  
criteria given above. Using the calculated cross section, this 
sample represents $\sim 22\pm4\%$ of all scanned events. The
average multiplicity of their charged particles is 
$\sim 1350\pm50$. From the multiplicity distribution per unit $\eta$, 
we get average particle
density\footnote[1]{The average multiplicity within $m$-th bin of
size $d\eta$ centered around the peak position, $\eta_{m}$, of
$dn/d\eta$-distribution.}, $<dn/d\eta>_{max}\approx 540\pm32$. This 
results, according to the frequently used Bjorken formula
\cite{bjork}, energy density of $3.743\pm0.324$
GeV/fm$^3$. This value is evidently larger than the density
required for the quark-gluon plasma formation \cite{satz-nature}.

\subsubsection{One-dimensional Analysis}
\label{sec2.1.1}

The pseudorapidity interval, \hbox{$2<\eta<6$},
is successively divided into $M$ equal bins.
The multiplicity in each such bins is counted and then the
corresponding $q$-order FM are calculated according to Eq.~\ref{Eq:1}.
In Fig.~\ref{fig:qgp1}A, FM are given in a log-log scale as functions of $M_{\phi}$
for the orders, \hbox{$q=\{2,3,4,5,6\}$}. The
underscore in $M_{\phi}$ refers to the patitioned
phase-space. The results of $\eta$-partition are given in Fig.~\ref{fig:qgp1}B. We notice that the
dependence of FM on $M_{\phi}$ shows almost the same characteristics
as in $\eta$-dimension. We also notice that the experimental
points can be fitted as straight lines. All lines have
positive {\it intermittency} exponents, $\phi_q$, which evidently
increase with increasing $q$. Generally, the
slopes in $\phi$- are larger than the slopes in
$\eta$-dimension.

\subsubsection{Two-dimensional Analysis}
\label{sec2.1.2}

The multi-dimensional FM ($\phi$, $\eta$,
$\cdots$) \cite{Ochs90} are suggested in order to study the interaction
dynamics, to study the sources of multiplicity fluctuations, 
and consequently to clearly explain the
power-law scaling. For the multi-dimensional partition, we should
first utilize either {\it isotropical} or {\it anisotropical} method. If
$\Delta\eta$, for example, is divided into the same number of
bins as $\Delta\phi$, this method called {\it self-affine}
partition (isotropical). Clearly, it leads to total partition
number, $M^2$. 
In Fig.~\ref{fig:qgp1}C, $F_q$ are drawn in dependence on
\hbox{$M_{\phi}\cdot M_{\eta}$} for the orders, \hbox{$q=\{2,3,4,5,6\}$}. 
In this picture, the 2D partition is performed isotropically. 
We notice that the experimental points can be fitted
as straight lines. All lines have positive slopes, 
$\phi_{\phi\eta}$. These slopes increase with increasing $q$. Obviously,
$\phi_{\phi\eta}$ are larger than the 1D ones 
($\phi_{\phi}$ and  $\phi_{\eta}$).

\subsection{Intermittency Phenomenon}
\label{sec2.2}

To study the intermittent behavior observed in Fig.~\ref{fig:qgp1}, we should
first check whether this phenomenon can be
understood by means of known physics, like conventional
{\it short-range correlations}. 
Elsewhere \cite{taw-jpc}, we discussed other interpretations, like
Bose-Einstein correlation (BEC), multi-particle cascade,
randomization, hadronic \'Cerenkov-radiation, measurement bias,
etc. It is known that the intermittent behavior is inconsistent 
with $\alpha$-, Lund hadronization, geometrical branching model, etc. 

\subsubsection{BEC and Coulomb Reactions}
\label{BEK-FM}

The contributions of BEC to the particle correlation have to
be considered, especially for identical particles. 
Furthermore, it is claimed that the intermittency is completely controlled by BEC 
and certain quantum statistical mechanism \cite{Chaos1,FM-Bec}. 
Including BEC in the FRITIOF Monte Carlo code, is was possible 
to simulate the like-charged two-particle correlations \cite{Na22}.  
In spite of these results, one should notice that the Dalitz decays and 
$\gamma$-conversations dominate the correlations, especially the 
lower-order ones of unlike-charged particles.
Using the emulsion technique, one has {\it almost} no chance to directly
determine neither the charges nor the momenta of produced
particles. These measurements are essentially required to 
estimate the BEC. Nevertheless, I have invented three different 
methods to estimate BEC in the emulsion \cite{TawDis}. It has been found that
BEC {\it nearly} dominate the interparticle correlations
especially within {\it very} small phase-space
intervals\footnote[2]{Within such small intervals, the
differences between the $\eta$-values of the selected particles are
correspondingly small. The relation between $\eta-$differences
and the relative momenta, $q^2=|\vec{q}_1-\vec{q}_2|^2$, is
given as follows \cite{TawDis}: \ba q^2&=&M^2-(im)^2\nonumber\\
&=&[q_{1t}^2+q_{2t}^2+2q_{1t}q_{2t}\cos(\phi_1-\phi_2)]+\nonumber\\
& &[M_{1t}^2+M_{2t}^2+2M_{1t}M_{2t}\cosh(\eta_1-\eta_2)]\nonumber\ea$M$
is the invariant mass.}, i.e. very small relative momentum, $q^2$. 
The intermittent behavior in {\it very} small
phase-space intervals has been discussed elsewhere \cite{taw-jpc,TawDis},
where not only a linear upwards trend was observed but a steep {\it
exponential} one. This exponential increase can partially be 
understood according to BEC. The 2D partition 
method represents another source for this exponential
upwards increasing \cite{emu012,EMU01b}. For relatively 
small $M$, as that in Fig.~\ref{fig:qgp1}, 
both of BEC and 2D partition method play a neglected 
role in FM \cite{taw-HEP2}.\\

The Coulomb reactions represent an additional effect on FM. 
These final state interactions take place, when the produced particles 
are closely emitted. Depending whether the particles are bosons or fermions, 
these reactions increase oder decrease the relative momenta, $Q$.  
Coulomb final state interactions have therefore, non-neglectable 
effects on BEC and then on FM, espetially for small $Q$.
The reactions between the identical poins and between the fragments represent 
about $10\%$ for $Q\le1~$MeV/c \cite{Gyula1}. For increasing $Q$, 
the reaction rate decreases, exponentially \cite{TawDis}. 
It can be neglectable for few MeV/c. 
Therefore, for small $M$, as in Fig.~\ref{fig:qgp1}, the contributions 
of Coulomb final state interaction can be neglected.

\subsubsection{Physics of Intermittent behavior}
\label{sec2.2.2}

Here, we introduce another physics
for the intermittent behavior. We will examine whether it is
able to describe the experimental data. In such as way, we simultaneously
estimate the responsibility of the critical transition on the
intermittent behavior, as introduced in Sect.~\ref{sec1}.

\begin{enumerate}
\item {\bf Self-similar processes} \cite{Bia86,Hwa-Self,Dahl} 
(e.g. QCD parton cascading). At the critical point of phase transition, the correlation 
length will be on the increase and under the scale transformation the fluctuations 
are expected to display self-similarity. If the gaussian approximation can 
be utilized to describe the particle production and from Eq.~\ref{Eq:3}, we get

\be
\frac{\phi_q}{\phi_2}=\frac{d_q}{d_2}\cdot(q-1).
\label{Eq:5}
\ee

As a particular case of this general description, we refer to the
mono-fractal behavior ({\it see} Eq.~\ref{Eq:8} below). The
implementation of L\'evy index $\mu$ \cite{Brax,Levy1,Levy2,Levy3} reads

\be
\frac{\phi_q}{\phi_2} = \frac{q^{\mu}-q}{2^{\mu}-2},
\label{Eq:6}
\ee

$\mu$ has a continuous spectrum within the {\it
region of stability}\footnote[3]{In different experiments, such as that in 
\cite{NA22,Heg}, it has been found  that the index, $\mu$, 
can be even outside this stability region.}, $\{0,2\}$. 
The index, $\mu$, allows an estimation of the cascading 
rate \cite{Brax}. Obviously, Eq.~\ref{Eq:5} cannot be not 
applied in the tails of these 
distributions. Whereas, Eq.~\ref{Eq:6} is more effective.
The two boundaries of L\'evy index, drawn in Fig.~\ref{fig:qgp3}D,
are corresponding to the degree of fluctuations in 
particle production as follows:
\begin{enumerate}
\item $\mu=2$, minimum fluctuations from self-similar
  branching processes. This is a suitable condition to apply
  the scaling rule \cite{Bia86},

\be
\frac{\phi_q}{\phi_2}=\left(\begin{array}{c} q \\ 2 \end{array}\right)
\label{Eq:7}
\ee

Therefore, from Eq.~\ref{Eq:5} the anomalous fractal ratios, $d_q/d_2$, 
are equal to $q/2$. For these conditions, ${\cal R}_q$ 
are corresponding to the multi-fractal processes. 
Later, we will realize that the last scaling rule is not able to describe
 the experimental results given in Fig.~\ref{fig:qgp3}A,B,C.
\item $\mu=0$, maximum fluctuations. Meanwhile
beneath the critical point and according to QCD, the correlation length 
 rapidly grows, it diverges suddenly according to the statistical quantum physics.
Therefore, near the critical point, 
i.e. neath the total randomization, it is expected that
$\{d_q|{\cal R}_q\}\rightarrow const.$ and $\phi_q\propto q-1$. 
At this point, the characteristics of 
 interacting system can be compared with that of 
 mono-fractal one \cite{Heg,Nazir}. This should not lead to the
conclusion that the mono-fractal behavior alone is sufficient for
the phase transition. In Fig.~\ref{fig:qgp3}, we will distinguish between
these quantities and ascertain the mono-fractal behavior.
\end{enumerate}

 Between these two limits, the approximation, Eq.~\ref{Eq:5}, is no longer
 valid and should be replaced by Eq.~\ref{Eq:6}.

\item {\bf Critical phase transitions} \cite{GQP2,Anton} 
(e.g. quark-hadron phase transition). If QGP indeed is to be produced, the
interacting system is supposed to suffer from thermal  phase transition
during its space-time evolution. At the critical point, \hbox{$\phi_q\propto q-1$}, 
\hbox{$d_q=d_2$}, and therefore, \hbox{${\cal R}_q/{\cal R}=1$}. 
In addition to the model introduced in Sect.~\ref{sec1}, 
there are auxiliary argumentations about utilizing the mono-fractality 
as signatures of the deconfinement.    
As the correlation length diverges at $T_c$ and if there are 
no long-range correlations, then $d_q$ are 
expected to be equal. Therefore, $d_q$ 
can be used to indicate the phase transition as follows: since the hadronization
is supposed to occur, only if the QGP-matter suffers from a phase transition, 
then the final hadron-matter is expected to be intermittent with constant $d_q$.
In the other case, if the hadronization
is ordinary a result of cascading processes, $d_q$ are linearly depending 
on $q$. Therefore, at the phase transition, the ratio of $q$-th intermittency slope to 
the second one is depending on the orders, $q-1$, 
(Eq.~\ref{Eq:3}),

\be
\frac{\phi_q}{\phi_2} = (q-1).
\label{Eq:8}
\ee

$\phi_2$ represents the dimension of the fractal sets in which the observed intermittent
behavior occurs \cite{Brax}. In Sect.~\ref{sec2.3.2}, we will get 
$\phi_2=1$ for vanishing $\mu$. Obviously, this value is an evidence 
for mono-fractality. These patterns are also expected during the
second-order phase transition from QGP to hadron gas \cite{Heg}. 
\end{enumerate}

Then, we conclude that the dependence of $\phi_q/\phi_2$ on $q-1$
reflects essential information about the reaction dynamics and about
the physics of intermittency patterns. 
Fig.~\ref{fig:qgp3} visualizes the differences between the {\it possible}
sources of intermittent phenomenon discussed above. Meanwhile 
Fig.~\ref{fig:qgp4} is used to encourage the conclusion, posted in this
letter. 

In next sections, the $q$-dependency of the
quantities, $\phi_q/\phi_2$, $d_q/d_2$, ${\cal R}_q/{\cal R}$, and
$\zeta_q$ will be examined and it will be shown 
how the intermittency and the fractal structure
of multiplicity fluctuations are used as signatures for 
the deconfinement phase transition \cite{GQP4}.

\subsection{Dependency on the orders of FM}

\setlength{\textfloatsep}{10pt}
\begin{figure}[htb]
\begin{center}
{ \epsfxsize=6cm \epsfbox{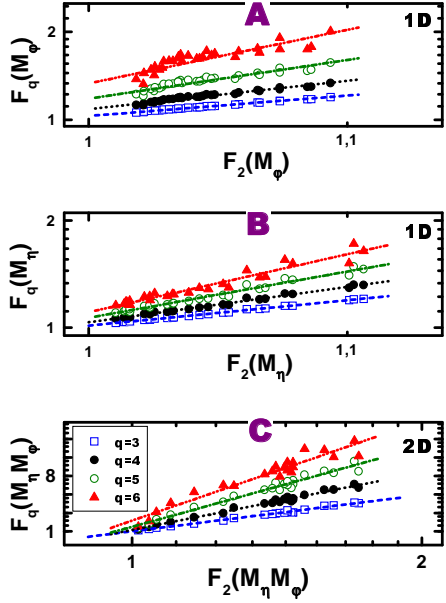} } 
 \parbox{9cm}{\caption{\small\it The relations between $q$-order FM
   and the second-order ones are drawn in log-log charts.
   The top picture, {\bf A}, depicts
   the analysis in $\phi-$dimension. The relations in $\eta-$dimension
   are given in picture {\bf B}. The bottom picture, {\bf C},
   illustrates the analysis in two-dimensions.
   In all pictures, the experimental points are linearly fitted. We notice that
   the slope ratios, $\phi_q/\phi_2$, increase with increasing $q$ 
	and they are in $\phi-$space larger than in $\eta-$space. Obviously, the 2D 
	slope ratios are larger than the 1D ones.  \label{fig:qgp2}
}}
\end{center}
\vspace{5pt} 
\end{figure}

In Fig.~\ref{fig:qgp1}, we noticed that the intermittency exponents,
$\phi_q$, are independent on $M_{\{\eta|\phi\}}$. 
Therefore, the ratios, $\phi_q/\phi_2$, can directly 
be deduced from the relation $F_q(M_{\{\eta|\phi\}})$ vs.
$F_2(M_{\{\eta|\phi\}})$,

\be
\log F_q(M_{\{\eta|\phi\}}) = \frac{\phi_q}{\phi_2} \log F_2(M_{\{\eta|\phi\}}) + c_q.
\label{Eq:9}
\ee

This power-law is also valid for the multi-dimensional partition 
method \cite{Fq-F2}. In a log-log scale, such relation
in $\phi$-dimension is depicted in Fig.~\ref{fig:qgp2}A 
for the orders, \hbox{$q=\{3,4,5,6\}$}. 
Fig.~\ref{fig:qgp2}B illustrates the
results in $\eta$-dimension. We notice that the
experimental points can be fitted as straight lines. Also, the
slopes clearly increase with increasing $q$. The results
from 2D analysis are given in Fig.~\ref{fig:qgp2}C. For the
comparison between $\phi_q/\phi_2$ in 1D and 2D,
one should first use anisotropical methods, 
in order to overcome the additional effects
of bin superposition especially in $\eta$-dimension 
\cite{taw-jpc,emu012,EMU01b}. 

\subsubsection{Dependency of $\phi_q/\phi_2$ on $q$}
\label{sec2.3.1}

Relation (\ref{Eq:9}) can be read as a power-law,

\be
F_q \propto F_2^{\beta_q}
\label{Eq:10}
\ee

In Fig.~\ref{fig:qgp2}, it is clear to recognize that the powers,
\hbox{$\beta_q\equiv\phi_q/\phi_2$}, increase with increasing $q$. 
Also the 1D $\beta_q$ in the rapidity dimension are smaller 
than that in the azimuthal space. The 
2D $\beta_q$ are apparently greater than the 1D ones. 

Using a
specific form of Ginzburg-Landau model to simulate the
deconfinement phase transition in heavy-ion collisions led to
the conclusion that $\beta_q$ are independent on the temperature
$T<T_c$. This guides the use of $\beta_q$ as a signature for the 
quark-hadron phase transition \cite{Nazir}.  

\be
\beta_q = (q-1)^{\nu}
\label{Eq:11}
\ee

The powers, $\nu$, describe the aftereffects of the phase
transition from chaotic to coherent state. In our case, the
aftereffects are the dimension ratios or basically the
fluctuations in the final state of particle production. 
Generally, $\nu$ are independent on the details of 
interacting system. Therefore, $\nu$ can be used to 
characterize the behavior of the measurable quantities at $T$ 
beneath $T_c$ (e.g. $\phi_q/\phi_2$, $d_q/d_2$, 
${\cal R}_q/{\cal R}$, $\zeta_q$, etc.). In this regard, $\nu$ are not a set of critical 
exponents in a conventional sense. 
Furthermore, they have a universal relevance \cite{Nazir}. \\

\setlength{\textfloatsep}{10pt}
\begin{figure}[htb]
\begin{center}{\epsfxsize=6cm \epsfbox{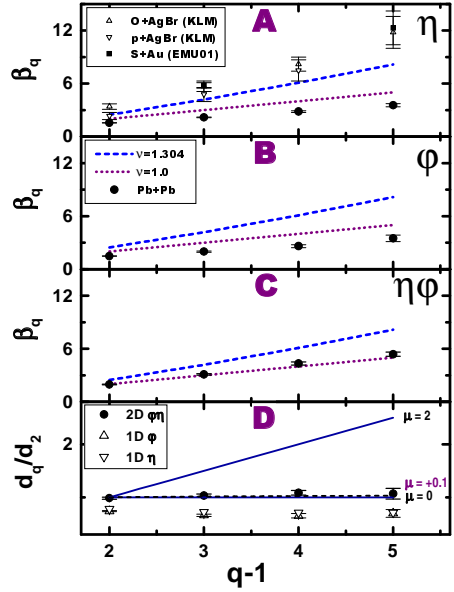}} 
 \parbox{9cm}{\caption{\small \it The dependences of the ratios,
\hbox{$\beta_q \equiv \phi_q/\phi_2$} on the orders,
\hbox{$q-1$}, for 1D are drawn in {\bf A} and {\bf B} for
$\phi-$ and $\eta-$dimension, respectively, where the
experimental results (solid points) cannot be predicted neither
by Eq.~\ref{Eq:8} nor by Eq.~\ref{Eq:11}. In {\bf A}, 
the triangles represent O-AgBr and p+AgBr at 200 AGeV \cite{KLMa}  
and the squares are for S+Au at 200 AGeV \cite{EMU01a}. 
Pb+Pb data are lower that the line with $\nu=1.0$, meanwhile 
the previous results are over the line with $\nu=1.304$. 
The 2D analysis is given in {\bf C}. $d_q/d_2$ are also 
studied in dependence on $q-1$ and drawn in {\bf D} (solid circles). 
The solid lines represent the two boundaries of
L\'evy stable law. The dash-line is the implementation of Eq.~\ref{Eq:6} 
for $\mu=+0.1$. The open triangles represent the 1D results. \label{fig:qgp3}
}}
\end{center}
\vspace{5pt} 
\end{figure}

In Fig.~\ref{fig:qgp3}A,B the relation  $\beta_q$ vs. $q-1$ 
are illustrated in $\eta$- and $\phi$-dimension, 
respectively. As discussed above, $\beta_q$ can directly be determined from the
slopes of the relation $\log F_q$ vs. $\log F_2$ (Eq.~\ref{Eq:9}). 
We notice that the experimental results cannot be predicted by
any of the lines given by Eq.~\ref{Eq:8} ($\nu=1.0$ \cite{satz1}) or
Eq.~\ref{Eq:11} ($\nu=1.304$ \cite{Nazir}). Comparing our results with previous
experiments of KLM- and EMU01-collaboration \cite{KLMa,EMU01a}, we notice that even though 
all results can not be fitted by any of these $\nu$-values, 
Pb+Pb results are relatively closer to the first line. 
Although, the other line is obtained 
from the Ginzburg-Landau description 
of a second-order phase transition \cite{Nazir}, 
it ist unable to fit any experimental data. 
As given in Sect.~\ref{sec2.2.2}, $\nu=1.0$ characterizes the 
critical phase transition. This value is also obtained from
2D Ising model  with second-order phase transition\footnote[4]{
On the one hand, the appearance of intermittent 
patterns during a phase transition has been proven in 2D Ising 
model long time ago \cite{satz21,satz1,Opall}. The intermittent fluctuations are 
characterized by a constant anomalous fractal dimension. 
On the other hand, we should utilize such statistical model to 
study the quark-hadron phase transition, since in lattice QCD 
at final temperature, which practically is close relevant to QGP, 
there is no direct calculation for the phase transition. Its characteristics can be 
obtained by numerical and analytical studies for statistical spin 
models.} \cite{satz1}. Ordinarily, $\nu=1.0$ refers to the 
mono-fractality. Experimentally, different values are  
obtained for $\nu$ \cite{GQP2,Pan}. But non of 
them can be related to the phase transition. 

We can conclude that the 1D results in either $\eta$- or 
$\phi$-dimension do not show clear signature for QGP. 
On the account of this restriction, one narrows
the QGP signatures either in the rapidity or in the location. 
This restriction ist that as it is the reason for this 
breakdown. For that reason, 
the search for QGP signatures should be 
performed through the posstible rapidity partitions and simulateneously 
over the possible locations. In such a way, we 
hope to disintegrate the fine scale of QGP.\\

The 2D dependence of $\beta_q$ on $q-1$ is 
given in Fig.~\ref{fig:qgp3}C. We notice that 
the relation (\ref{Eq:7}), which has been suggested 
in \cite{Bia86} and successfully utilized 
in \cite{EMU92}, is not able to fit the experimental data.
This means that the self-similar branching 
processes and consequently the minimal fluctuations are not
responsible for the power-law scaling (Sect.~\ref{sec2.2.2}). The argumentations given 
in Sect.~\ref{BEK-FM} are valid, if the intermittency is in fact the self-similar behavior. 
The dash-lines represent the power-fit according to 
Eq.~\ref{Eq:11} with \hbox{$\nu=1.304$} \cite{Nazir}. The same process, but
with \hbox{$\nu=1.0$}, is represented by the dotted lines (Eq.~\ref{Eq:8}). We
notice that the first power-fit is completely unable to describe the
experimental data. Meanwhile, the second one gives a line very close
to the experimental data (solid points). 
As given above, this result can be regarded as a 
signature of deconfinement phase transition. 
On the contrary to 1D results, we get $\nu=1.0$ for the 2D dependency 
of $\beta_q$ on $q$.\\

Furthermore, it is important to recognize that Eq.~\ref{Eq:8} and Eq.~\ref{Eq:11} 
are not belonging to L\'evy stable region. In the next section, 
we use L\'evy space to draw the $q$-dependency of $d_q/d_2$. 
In such a way, we try to reconfirm the previous results.

\subsubsection{Dependency of $d_q/d_2$ on $q$}
\label{sec2.3.2}

From Eq.~\ref{Eq:3} and Eq.~\ref{Eq:11} we get

\be
\frac{d_q}{d_2} = (q-1)^{\nu-1}
\label{Eq:12} 
\ee 

Fig.~\ref{fig:qgp3}D  illustrates this relation. 
We notice that the 1D results (open triangles) are completely 
outside the L\'evy stable region (Sect.~\ref{sec2.2.2}). In
spite of this result, it is a significant finding to realize,
that the values of $d_q/d_2$ are apparently independent on 
$q$ ($d_q/d_2\approx 0.66\pm 0.04$). Eq.~\ref{Eq:6} with
\hbox{$\mu=+0.1$}, is only able to describe
the 2D results (solid circles). The values of 2D $d_q/d_2$ are also 
constant ($d_q/d_2\approx 1.0\pm 0.03$). 

For small but distinctly non-zero $\mu$, the possibility of 
QGP formation mixed within cascading processes can 
not be withdrawn \cite{Brax}. If $\mu=0$, then 
$\phi_q/\phi_2=q-1$ and correspondingly $d_q/d_2=1$ 
(Eq.~\ref{Eq:3} and Eq.~\ref{Eq:6}). The predictions of 
Eq.~\ref{Eq:8} (\hbox{$d_q/d_2=1$}), 
which, as given above, emphatically  supports 
the conclusion of phase transition \cite{GQP2}, are at the
low boundary of L\'evy space. 
Once again, the obtained values of $d_q/d_2$ are not consistent with being
proportional to $q$ as claimed in \cite{EMU92}. 
On the contrary, they are equal.

\subsubsection{Dependency of ${\cal R}_q/{\cal R}$ on $q$}
\label{sec2.3.3}

\setlength{\textfloatsep}{10pt}
\begin{figure}[htb]
\begin{center}{\epsfxsize=6.5cm \epsfbox{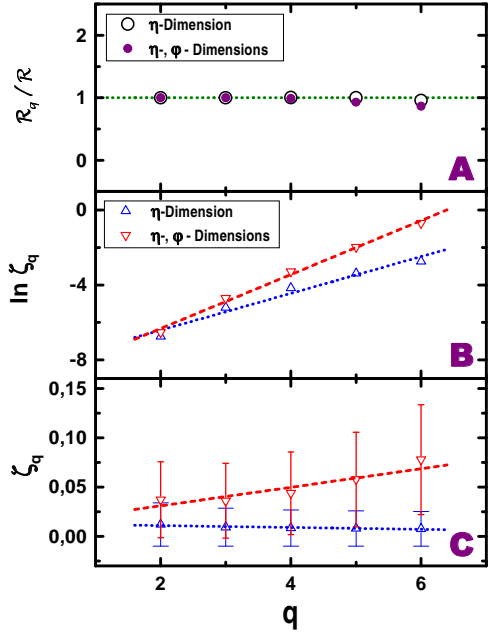}}  
\parbox{9cm}{\caption{\small\it In {\bf A} fractal R\'enyi dimensions,
${\cal R}_q/{\cal R}$, are depicted versus $q$. 
For 1D as well as for 2D data, we notice that \hbox{${\cal
R}_q/{\cal R}\approx 1$}. In 
{\bf B}, the relations between \hbox{$\log \zeta_q$} and $q$ in 1D and 2D are
illustrated. In contrast to \cite{EMU92} we have here a clear
exponential dependence of $\zeta_q$ on $q$. 
In the bottom picture, {\bf C}, $\zeta_q$ are tested as function of 
the order, $q$, according to Eq.~\ref{Eq:14}. The linear fits are implied by the lines.
\label{fig:qgp4}
}}
\end{center}
\vspace{5pt} 
\end{figure}

As discussed above, ${\cal R}_q$, 
one the one hand, measure the degree of randomization in the final 
state of particle production. On the other hand, they 
represent the fractal critical dimensions of the randomized 
intermittent clusters observed in the confined matter which 
is entirely bedded inside the deconfined one. 
As given in Sect.~\ref{sec1}, the thermal phase transition can be 
marked by constant ${\cal R}_q$. The $q$-dependency of 
${\cal R}_q$ in 1D and 2D is depicted in Fig.~\ref{fig:qgp4}A. 
As expected, we notice that
\hbox{${\cal R}_q/{\cal R}\approx 1$}. Near the critical point,
\hbox{$d_q/d_ 2\rightarrow 1$}. Then the corresponding R\'enyi
dimensions, ${\cal R}_q\rightarrow {\cal R}$, and 
$\phi_q/\phi_2\propto q-1$. 

As given in Sect.~\ref{sec2.3.4} below, near the critical point, 
the normalized exponents, $\zeta_q$, show an
abnormal behavior. They have an exponential $q$-dependency 
rather than a linear one. This will be utilized to distinguish between the
scaling rules given in Sect.~\ref{sec2.2.2}.

\subsubsection{Dependency of $\zeta_q$ on $q$}
\label{sec2.3.4}

The scaling rule, Eq.~\ref{Eq:7}, is only valid, if the correlated
$q$-tuplets are built up from single correlated particle-pairs
and \hbox{$q-2$} non-correlated ones. Starting from the
assumption given in \cite{EMU92} that FM are defined as the 
integration of multi-particle
correlation functions together with the lower-order ones which
strongly effect the higher-order ones and defining the
normalized exponents as

\be
\zeta_q\equiv \phi_q \left/ \left(\begin{array}{c} q \\  2 \end{array}\right)\right.,
\label{Eq:13} 
\ee

it is assumed that $\zeta_q$ is linearly depending on $q$ (as given in Fig.6 of Ref.~\cite{EMU92}). 
In Fig.~\ref{fig:qgp4}B, we check this relation for 1D as well as for 2D FM.
It is clear to notice that $\zeta_q$ exponentially increase with increasing
$q$. This result supports, on the one hand, the
assumption that Eq.~\ref{Eq:7} is not able to describe the ratios
$\phi_q/\phi_2$. On the other hand, it supports the conclusion that $d_q$ are
constant for all orders of FM, as we noticed in Fig.~\ref{fig:qgp3}. \\

Using the scaling-rule, Eq.~\ref{Eq:8}, to define the normalized exponents, we get 

\be
\zeta_q \equiv \phi_q \left/ (q-1)\right.,
\label{Eq:14} 
\ee

Eq.~\ref{Eq:14} has been tested in Fig.~\ref{fig:qgp4}C.  
We find that $\zeta_q$ are linearly depending on $q$. This guides  
to the conclusion that meanwhile Eq.~\ref{Eq:7} failed 
to describe the Pb+Pb data, Eq.~\ref{Eq:8} is obviously able. 
Otherwise, we notice that $\zeta_q$ in 1D decrease 
with increasing $q$, whereas in 2D there is a positive increasing. 
 
\section{Summary and Conclusions}
\label{sec3}

In this latter, the non-statistical fluctuations
in Pb+Pb collisions at 158 AGeV are investigated. First, I would like to sum up
the results obtained so far. The analysis of FM in 1D and 2D
shows that this interacting system is obviously
intermittent. The values of 1D $\beta_q$ are considerably lower than 
the results obtained by KLM- and EMU01-collaboration.  
Eq.~\ref{Eq:7} suggested in \cite{Bia86} and utilized in \cite{EMU92} 
faild to simulate the dependency of 1D and 2D $\beta_q$ on  
\hbox{$q-1$}. Meanwhile, Eq.~\ref{Eq:8} results a  
line in 1D more close to Pb+Pb than to the other data, the results of 2D $\beta_q$ 
can effectively be fitted by this equation. The behavior described 
by Eq.~\ref{Eq:8} is predicted, on the one hand, if the interacting 
system suffers from thermal phase transition during its space-time 
evolution. On the other hand, the disability of this equation to perfectly 
describe the 1D ratios leads to the conclusion that the
Pb+Pb data are not uniquely mono-fractal. This results should not 
disturb the main conclusion posted in this letter, since on the basis of 
the fine scale of QGP, one can expect, that its signatures might be locally oriented. 
Therefore, the individual analysis in rapidity- or azimuthal-space may 
not be favorable to detect QGP. The 2D analysis is therefore, more effective than the 1D.
In addition to these results, we get \hbox{$d_q/d_2=1$} in 2D 
investigation of the anomalous fractal dimensions. 
This value confessedly supports the 
hypothesis of the deconfinement phase transition. L\'evy index, 
\hbox{$\mu=0.1$}, obtained for this 2D analysis, is an additional 
confirmation of this transition. We noticed that the fractal R\'enyi 
dimensions are constant (\hbox{${\cal R}_q={\cal R}$}).  
Also, in Eq.~\ref{Eq:13} the relation between the normalized exponents, 
$\zeta_q$, and the orders of FM shows that $\zeta_q$
exponentially increase. This dependence assists the conclusion
that the relation (\ref{Eq:7}) is not able to describe the experimental data 
and therefore, it should be replaced by Eq.~\ref{Eq:8}, 
as done in Eq.~\ref{Eq:14}. \\

Finally, we come to the conclusion that the data sample used in
this letter is intermittent in 1D as well as in 2D.
The intermittency ratios can be given by using simple relations,
as that given in Eq.~\ref{Eq:8}. Furthermore, 
according to certain statistical models, the $q$-dependency of  
anomalous fractal and R\'enyi dimensions and the index 
$\mu$ evidently support the deconfinement phase 
transition. More data are however, needed to confirm the 
results and interpretations reported in this letter. 

\section*{Acknowledgements}
I am are very grateful to E.Stenlund and all colleagues of 
the EMU01 collaboration for the kind assistance. 
Especially that they allowed the use of part of our 
collaborative experimental data for the present work. 
Also I would like to thank E.Ganssauge, and
F.P{\"u}hlhofer for the helpful discussions 
and the contentious support.



\begin{thebibliography}{99}

\bibitem{GQP1} L.van Hove, {\it Z.Phys.} {\bf C21} 93 (1983) \\
           	M.Gyulassy, {\it et al.}, {\it Nucl.Phys.} {\bf B237} 477 (1984) \\
           	M.Gyulassy, {\it Nucl.Phys.} {\bf A418} 59c (1984)

\bibitem{GQP2} A.Bia{\l}as, R.C.Hwa, {\it Phys.Lett.} {\bf B253} 436 (1991)

\bibitem{GQP3} R.C.Hwa, {\it Nucl.Phys.} {\bf A525} 537c (1991)

\bibitem{GQP4} E.Shuryak, {\it Phys.Lett.} {\bf B423} 9 (1998) \\
			G.Baym, H.Heiselberg, {\it Phys.Lett.} {\bf B469} 7 (1999) \\
			M.Asakawa, U.Heinz, B.M\"uller, {\it Phys.Rev.Lett.} {\bf 85} 2072 (2000)

\bibitem{jacee} T.Burnett {\it et al.}, {\it Phys.Rev.Lett.} {\bf 50} 2062 (1983)

\bibitem{spik} M.Adamus, {\it et al.}, {\it Phys.Lett.} {\bf B185} 200 (1987)

\bibitem{Bia86} A.Bia{\l}as, R.Peschanski, {\it Nucl.Phys.} {\bf B273} 703 (1986),
           	{\it Nucl.Phys.} {\bf B308} 857 (1988)

\bibitem{Rajag} B.Berdnikov, K.Rajagopal, {\it Phys.Rev.} {\bf D61} 105017 (2000)

\bibitem{taw12} A.M.Tawfik, E.Ganssauge, {\it Nucl.Instr.Methods} {\bf A416} 136 (1998)\\
           	A.M.Tawfik, {\it Comput.Phys.Commun.} {\bf 118}  49 (1999)

\bibitem{taw-jpc} A.M.Tawfik, E.Ganssauge, {\it Heavy Ion Phys.} {\bf 12} 53 (2000).

\bibitem{EMU01-1} M.I.Adamovich, {\it et al.}, {\it Phys.Lett.} {\bf B390} 445 (1997)

\bibitem{TawDis} A.M.Tawfik, {\it ''Kritische Studien zu der
        	Teilchenkorrelation und den Signaturen des Phasen\"ubergangs''}, 
			(Tectum-Verlag, Marburg, 1999)  [in German]

\bibitem{bjork} J.D.Bjorken, {\it Phys.Rev.} {\bf D27} 140 (1983)

\bibitem{Hwa-Self} R.C.Hwa, {\it Quark-gluon plasma}, ed. R.C.Hwa, 
			(World Scientific, Singapure, 1990) 

\bibitem{Chaos1} P.Carruthers, {\it at al.}, {\it Phys.Lett.} {\bf B222} 487(1989)

\bibitem{Chaos2} B.-L.Hao, {\it Chaos}, (World Scientific, Singapure, 1984) \\
    		P.Berg\'e, Y.Pomeau, C.Vidal, {\it Order within Chaos}, (Wiley, New York, 1984)   

\bibitem{Bial11} W.Kittel, R.Peschanski, European Physical Society Conf., Madrid, 
			Spain, Sep 6-13, 1989
\bibitem{Bial12} 	R.Peschanski, Workshop on Large-Scale Structures in Non-Linear Physics, 
			Villefranche-Sur-Mer, France, Jan 13-18, 1991 
\bibitem{Bial13} 	W.Ochs, {\it Phys.Bl.} {\bf 46} 123 (1990) 

\bibitem{satz21} J.Wosiek, {\it Acta Phys. Polon} {\bf B19} 863 (1988)   

\bibitem{satz22} B.Bambah, H.Satz, {\it Nucl.Phys.} {\bf B332} 629 (1990)

\bibitem{satz1} H.Satz, {\it Nucl.Phys.} {\bf B326} 613 (1989)
 
\bibitem{Hent} H.G.Hentschel, I.Procaccia, {\it Physica}, {\bf D8} 435 (1983)

\bibitem{Reny1} R.Peschanski, {\it Nucl.Phys.} {\bf B235} 317 (1990)

\bibitem{Reny2} P.Lipa, B.Buschbeck, {\it Nucl.Lett.} {\bf B223} 465 (1989)

\bibitem{Opall} R.Peschanski, {\it L\'eon van Hove Festschrift} 
			eds. A.Giovannini, W.Kittel, World Scientific 1990 

\bibitem{Reny3} S.Gupta, P.La Cock, H. Satz, {\it Nucl.Phys.} {\bf B362} 583 (1991)

\bibitem{Cross-S} L.Y.Geer, {\it et al.}, {\it Phys.Rev.} {\bf C52} 334 (1995) \\
			B.S.Nilsen, {\it et al.}, {\it Phys.Rev.} {\bf C52} 3277 (1995) 

\bibitem{evert} E.Stenlund, {\it Nucl.Phys.} {\bf A590} 597c (1995)

\bibitem{satz-nature} H.Satz, {\it Nature} {\bf 324} 116 (1986)

\bibitem{Ochs90} W.Ochs, {\it Phys.Lett.} {\bf B247} 101 (1990)

\bibitem{FM-Bec} M.Charlet, Sov.J.Nucl.Fiz. {\bf 56} 1497 (1993)

\bibitem{Na22} N.Agababyan, {\it et al.}, {\it Phys.Lett.} {\bf B332} 458 (1994) \\
			T.Wibig, {\it Phys.Rev.} {\bf D53} 3586 (1996)

\bibitem{emu012} M.I.Adamovich {\it et al.}, {\it Z. Phys.} {\bf C76} 659 (1997).

\bibitem{EMU01b} W.Yuanfang, {\it et al.}, {\it Phys.Rev.Lett.} {\bf 70} 3197 (1993)\\ 
			L. Lianshou, {\it et al.}, {\it Phys.Lett.} {\bf B388} 10 (1996); 
			{\it Z.Phys.} {\bf C69} 323 (1996); {\it Z.Phys.} {\bf C73} 535 (1997)

\bibitem{taw-HEP2} A.M.Tawfik, to be appear in {\it Heavy Ion Phys.}, \hbox{hep-ph/0012022} 

\bibitem{Gyula1} M.Gyulassy, S.Kaufmann, {\it Nucl.Phys.}, 503 {\bf B362}
 
\bibitem{Dahl} P.Dahlqvist, B.Andersson, G.Gustafson, {\it Nucl. Phys.} {\bf B328} 76 (1989)  

\bibitem{Brax} Ph.Brax, R.Peschanski, {\it Phys.Lett.} {\bf B253} 225 (1991)

\bibitem{Levy1} W.Ochs, {\it Z. Phys.} {\bf C50} 339 (1991) 

\bibitem{Levy2} Y.Zhang, L.Liu, Y.Wu, {\it Z.Phys.} {\bf C71} 499 (1996) 

\bibitem{Levy3} B.W.Gnedenko, A.N.Kolmogorov, {\it ''Grenzverteilungen von Summen
         	unanbh{\"a}ngiger Zufallsgr{\"o\ss}en''} (Akademie-Verlag, Berlin, 1960) [in German]

\bibitem{NA22} N.Agababyan, {\it et al.}, {\it Z.Phys.} {\bf C59} 405 (1993) 

\bibitem{Heg} S.Hegyi, {\it Phys.Lett.} {\bf B318} 642 (1993)

\bibitem{Nazir} R.C.Hwa, M.Nazirov, {\it Phys.Rev.Lett.} {\bf 65} 741 (1992)

\bibitem{Anton} N.G.Antoniou, {\it et al.}, {\it Phys.Rev.} {\bf D45} 4034 (1992)

\bibitem{Fq-F2} W.Ochs, J.Wosiek, {\it Phys.Lett.} {\bf B214} 617 (1988) \\ 
          	R.C.Hwa, {\it Phys.Rev.} {\bf D47} 2773 (1993)

\bibitem{KLMa} R.Holynski, {\it at al.}, {\it Phys.Rev.Lett.} {\bf 62} 733 (1989)

\bibitem{EMU01a} M.I.Adamovich, {\it at al.}, {\it Phys.Rev.Lett.} {\bf 65} 412 (1990)

\bibitem{Pan} J.Pan, R.C.Hwa, {\it Phys.Rev.} {\bf D46} 4890 (1992) 

\bibitem{EMU92} M.I.Adamovich, {\it et al.}, {\it Nucl.Phys.} {\bf B388} 3 (1992)
\end{thebibliography}
\end{document}